\documentclass{WileyMSP-template}

\usepackage{xcolor}
\usepackage{amsmath}
\usepackage{amssymb}
\usepackage{yfonts}
\usepackage{array}
\renewcommand{\vec}[1]{\mathbf{\boldsymbol{#1}}}
\newcommand{\Cal}[1]{\mathcal{#1}}
\newcommand{\Goth}[1]{\mathfrak{#1}}

\definecolor{mygreen}{RGB}{138, 182, 7}
\definecolor{mybrown}{RGB}{204, 153, 0}

\begin{document}

\pagestyle{fancy}
\rhead{\includegraphics[width=2.5cm]{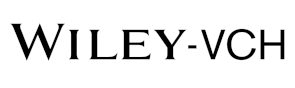}}

\title{On flat band engineering and graphene nanomesh magnonics
}

\maketitle


\author{Vasil Saroka*}


\dedication{}

\begin{affiliations}
V.~A.~Saroka\\
$^{1}$Department of Physics, University of Rome Tor Vergata and INFN, Via della Ricerca Scientifica 1, 00133 Roma, Italy\\
$^{2}$Institute for Nuclear Problems, Belarusian State University, Bobruiskaya 11, 220006 Minsk, Belarus\\
Email Address: vasil.saroka@roma2.infn.it
\end{affiliations}



\begin{abstract}
Graph-theoretic topological frustration is a peculiar situation on a finite piece of the honeycomb lattice that prevents a full pairwise coupling of the lattice sites via nearest neighbor links, even when the total number of sites is an even number and the structure possesses no sublattice imbalance. This type of frustration is inherent for organic molecules that are classified as concealed non-Kekulean hydrocarbons, representing peculiar diradicals. Here we show that this topological frustration persists in two-dimensional systems based on the honeycomb lattice. Such systems exhibit fully flat electronic energy bands located at the Fermi level. Therefore, two-dimensional ultimately flat bands can be systematically and predictably constructed for graphene monolayer nanomeshes. These systems are prone to antiferromagnetic ordering and hybrid spin-wave excitations mixing weak ferromagnetic and strong antiferromagnetic features, which could pave the way towards low-power, compact, and ultrafast organic spintronics with near room-temperature operation.
\end{abstract}


\section{Introduction}
Flat bands are key for the emergence of strongly correlated physics, wherein quasiparticles interactions are stronger than typical kinetic energy given by the energy band width~\cite{Georges1996,Lee2006,BookAmusia2020}. 
The high density of states associated with narrow band width can result in Hubbard-Mott-Stoner magnetism~\cite{Stoner1938,Stoner1939,Hubbard1963,Qaiumzadeh2009,Jing2019,Ortiz2023,Yu2024}, unconventional and high-temperature superconductity~\cite{Khodel1990,Volovik1994,Shaginyan2010,Kopnin2011,Kopnin2013,Esquinazi2014,Volovik2018,Peotta2015,Shaginyan2022,Shaginyan2022b,Tian2023,Han2025}, and an anomalous fractional quantum Hall effect~\cite{Tang2011,Neupert2011,Wang2011,Sheng2011,Regnault2011,Shi2025}.
Although some iconic flat band lattices, such as dice~\cite{Sutherland1986}, Lieb~\cite{Lieb1989}, and Kagome~\cite{Mielke1992,Hanisch1995} lattices, are known for a long time, the general rules for flat band construction are somewhat obscure and so far deal with very exotic examples and/or incomplete flatness~\cite{SuarezMorell2010,Green2010,MoralesInostroza2016,Heikkila2016,Ramachandran2017,Pal2018,Pal2018b,Kolovsky2018,Mao2020,Morfonios2021,Kim2023b,Zheng2025}. Significant numerical efforts have been undertaken to reveal potential candidates with band flatness~\cite{Regnault2022}.
Experimentally, major success has been achieved with twisted layers of the honeycomb lattice~\cite{Cao2018a,Cao2018b,Cao2020,Park2021a,Boi2024} and Kagome lattice materials~\cite{Ye2018,Ortiz2019b,Kang2020,Tan2021,Kang2022,Kim2023,Kassem2024}, including their supramolecular analogs~\cite{Pawlak2025,Yan2026}. Recent work also reports progress towards solid state realization of the Lieb lattice~\cite{Devarakonda2025,Li2026a}. A somewhat broader success in flat band design and control has been achieved with artificial systems, for example, in photonic metamaterials and optically trapped cold atoms. The latter includes dice lattice realization~\cite{Bercioux2009}, Lieb lattice manipulation~\cite{Whittaker2018,Whittaker2021a}, light-matter interaction enhancement in free-electron lasers~\cite{Yang2023}, and fascinating orbital rotation in Kagome lattice~\cite{Xia2025}.

Although three-dimensional (3D)~\cite{Nishino2005,Bergman2008,Guo2009} and one-dimensional (1D)~\cite{Tasaki1995,Tasaki1996,Kusakabe2003,Kohmoto2007,Tasaki2008} flat bands can be designed, nowadays the main interest is around two-dimensional (2D) bands due to the abundance of existing~\cite{Miro2014,Ren2025} and newly predicted~\cite{Meng2024} 2D materials, which in a broader context are related to the practical importance of surface science and planar lithographic technology. It is important to note that there are few fundamental theorems that prohibit long-range orders in 1D and 2D systems: the Lieb-Mattis theorem for ferromagnets~\cite{Lieb1962}, the Anderson theorem for antiferromagnets~\cite{Anderson1952} and the Hohenberg-Mermin-Wagner theorem covering superconductors and anti-/ferro-magnets~\cite{Mermin1966,Hohenberg1967}. This seems to undermine any flat band considerations beyond 3D systems. Yet, recent theoretical revisions imply that finite range orders may be macroscopically large enough in 2D~\cite{Palle2021}; alternatively, the restrictive conditions --- short range interactions and continuous symmetries~\cite{Mermin1966,Hohenberg1967} --- can be bypassed by material engineering, which is in agreement with the existence of twisted bilayers~\cite{Cao2018a,Cao2020,Cheung2021} and atomically thin ferromagnets: CrI$_3$~\cite{Huang2017} and~Cr$_2$Ge$_2$Te$_6$~\cite{Gong2017}. These peculiarities make the 2D flat bands extremely compelling to study.

In general, 2D flat bands can be located at any energy of the electronic spectrum. They can also have intersections with other dispersive bands. Unless the latter is a special demand, the intersections must be avoided, so that flat bands are well isolated from the remaining spectrum, thereby enabling unambiguous detection and interpretation of the associated phenomena. In the Kagome lattice, depending on the parameters of the model, the flat band is either the lowest or the highest in energy, while in the dice and Lieb lattices such a band is pinned to the Fermi level. For bosonic systems, which do not follow Pauli's exclusion principle, the flat band shall be the lowest in energy~\cite{Wu2007}. The whole-spectrum flatness may be needed to ensure a uniform response across a wide energy range of excitations~\cite{Xia2025}.
However, for fermionic systems, the most attractive flat bands are those positioned at the borderline between occupied and empty states, which is referred to as the Fermi energy and commonly is set as a reference zero energy.

In the area of organic chemistry, the zero-energy states of non-Kekulean hydrocarbons~\cite{Gutman1974,Cyvin1989,Cyvin1990,Ortiz2019,Mishra2020a} are described by the graph-theoretic situation called topological frustration~\cite{Wang2008,Wang2009a,Bullard2014}. It is important to note that this frustration is physically different from the one studied in frustrated magnets and spin glasses~\cite{Diep2025}, for example in an antiferromagnet with a triangular or Kagome lattice. In contrast to negatively coupled spins occupying \emph{odd} number of lattice sites in the frustrated antiferromagnet, topological frustration in hydrocarbons deals with electron pairing in covalent bonds. This electron pairing is the one from resonating valence bond states, which inspired quantum spin liquids~\cite{Anderson1973,Savary2017}. Full pairing between odd number of sites is obviously impossible. Full pairing between \emph{even} number of sites is trickier, especially when dealing with bipartite lattices such as the honeycomb lattice of hydrocarbons. When a there is an imbalance between number of atomic sites in the two sublattices of a bipartite lattice, the pairing is called \emph{class~I topological frustration}~\cite{Wang2008,Wang2009a}, while the hydrocarbon molecules exhibiting it are called \emph{obvious non-Kekuleans}~\cite{Cyvin1989}). Class~I topological frustration allows generic design of zero-energy states, which can be translated into flat bands for periodic structures, see Ref.~\cite{Calugaru2022} and literature therein. This approach is used, for instance, in triangulene-based covalent organic frameworks~\cite{Ortiz2023,Catarina2023,Phillips2026}. However, another peculiar situation is possible with respect to electron pairing. Even with zero sublattice imbalance, the full pairing on the honeycomb lattice can be impossible. This special case is called \emph{class~II topological frustration}~\cite{Wang2009a}. Hydrocarbon molecules featuring it are dubbed \emph{concealed non-Kekuleans}~\cite{Cyvin1989}. 
Later in the main text, by \emph{topological frustration}, we will mean only class~II topological frustration because it is so peculiar and subtle that until quite recently, it was not even clear if class~II topological frustration can be translated into periodic structures. In case of success, this would automatically supply a new approach to flat band engineering.
At the mathematical level, all above frustrations are connected to the matching theory of graphs~\cite{BookLovasz2009}. 
Recently, using matching theory, it has been shown that class~II topological frustration is not limited to molecules. Quasi-1D polymers can exhibit it, too~\cite{Saroka2026b}. In fact, nothing prevents the existence of the topological frustration in 2D systems either as we shall demonstrate later.

2D flat bands originating from topological frustration of class~II on honeycomb lattice, would be of useful service because zero sublattice imbalance together with high density of states shall lead to antiferromagnetic (AFM) spin ordering~\cite{Lieb1989,Ovchinnikov1978}. On one hand, the AFM order has practical benefits because it can lead to more compact spintronic devices due to the net zero magnetization and lack of stray fields as compared to ferromagnetic (FM) phase~\cite{Baltz2018,Siddiqui2020}. On the other hand, AFM phase is of fundamental interest, because it usually accompanies high temperature superconductivity~\cite{Lee2006,Cao2018a,Park2021a}. 

Generally speaking, the power of topological frustration comes form the fact that it amends the Hubbard-Stoner's criterion: 
\begin{align*}
    \text{flat band/high density of states} & \rightarrow \text{magnetism} \, .
\end{align*}
By attaching class~I and~II topological frustrations, one can gain predictive power with respect to the magnetic order type: 
\begin{align*}
    \text{topological frustration class~I} &\rightarrow \text{flat band} \rightarrow \text{FM phase}\, , \\
    \text{topological frustration class~II} & \rightarrow  \text{flat band} \rightarrow \text{AFM phase}\, ,
\end{align*}
which defines dispersions of magnetic excitations -- magnons -- available in the system for spintronic applications.
Flat bands ensure that magnetic order persists till vanishing electron-electron interactions.

In this paper, we show how 2D flat bands can be constructed in a systematic way on a familiar honeycomb lattice with the aid of class~II topological frustration considered on a torus instead of a plane, as is usual for hydrocarbons and finite flakes of nanographene~\cite{Cyvin1989,Cyvin1990,Wang2008,Wang2009a,Bullard2014}, or a cylinder, as recently proposed for polymers~\cite{Saroka2026b}. Flat bands are demonstrated in the nearest neighbor tight-binding (TB) formalism and 
then the Hubbard mean-field (TB+U) and linear Heisenberg models are used for investigation of the magnetic properties arising from the designed flat bands.

\section{Results and discussion}
\subsection{Graph theory settings\label{sec:GraphTheory}}
Graphs are collections of points called vertices, which are connected by lines called edges~\cite{BookOmelchenko2018}. Matchings are disjoint pairs of connected points. Topological frustration (class~II) is the property of a graph which prevents a \emph{perfect pairing} between the points, such that no unpaired points are left. Maximizing the number of matched points -- matchings -- and minimizing the number of free points, called the \emph{deficit} $\eta$, is one of the important graph theory problems~\cite{BookLovasz2009}. A brute force combinatorial approach to this problem leads one to an exponential time scalings, but polynomial time algorithms are possible: the Hopcroft-Karp-Karzanov algorithm for bipartite graphs~\cite{Hopcroft1973,Karzanov1973} and blossom algorithms for generalist graphs~\cite{Edmonds1965,Kolmogorov2009}.

It is possible to classify and study graphs by their allowed embeddings into orientable surfaces~\cite{Kasteleyn1967}: plane, cylinder, torus etc. Such embedding are useful mathematically and also establish direct relations to physical crystallographic structure~\cite{Girao2023,Macmillan2026}. Mathematically, it is interesting if some structural properties of the graphs persist when moving from one surface to another and how such properties can be preserved when graphs are updated while moving from one surface to another.

\begin{figure}
\centering
  \includegraphics[width=0.80\linewidth]{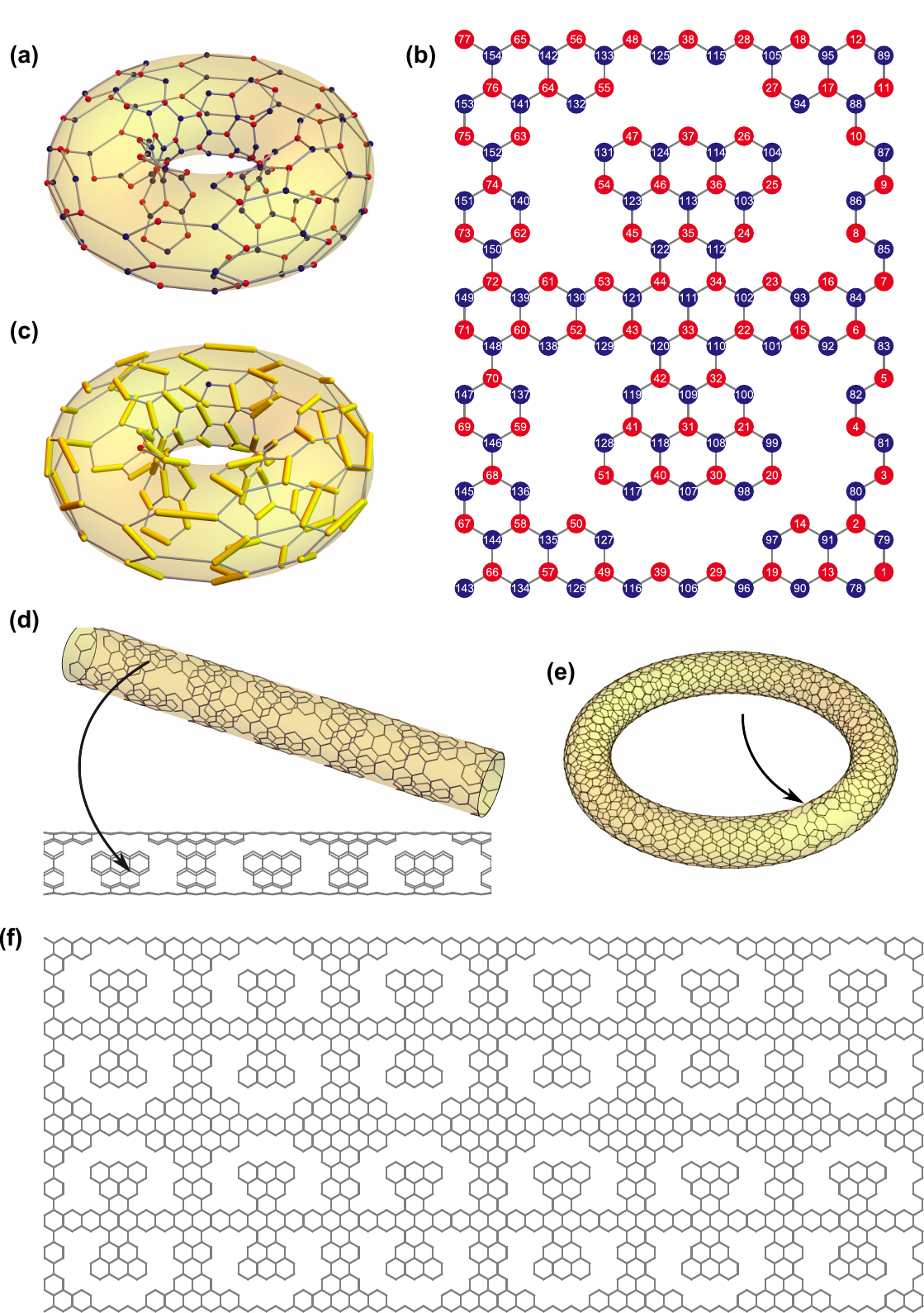}
  \caption{Topological frustration on torus. (a)  A torus embedment of bipartite hexagonal graph, featuring equal number of vertices in the red and blue sublattices. (b) The graph from panel (a) unrolled into a plane with explicit numbering of vertices shown. Edges connecting bottom to top and left to the right are not shown for clarity of the picture. (c) The maximal matching (yellow) returned by the Hopcorft-Karp-Karzanof algorithm. The two non-covered vertices (red and blue) represent the graph deficit $\eta=2$. (d) A segment of an amchair carbon nanotube $(6,6)$, decorated with nanocarving equivalent to the pattern in panel (b). A 2D side projection is shown below. Black arrow: view direction. (e) A torus based on long $(6,6)$ nanotube with an embedment of a single unit from panel (b) (black arrow). (f) A piece of the 2D topologically frustrated GNM, which unit cell is the graph in panel (b).}
  \label{fig:TopologicalFrustrationOnTorus}
\end{figure}

In Figure~\ref{fig:TopologicalFrustrationOnTorus}a, b we construct a graph on the torus corresponding to a doubly periodic structure, that possess the same topological frustration property as the celebrated finite size molecule called Clar's goblet, synthesized recently with bottom-up surface assisted method~\cite{Mishra2020a}.
The absence of perfect matching is directly verified by the Hopcroft-Karp-Karzanov algorithm, which returns the graph deficit $\eta = 2$, as seen in Figure~\ref{fig:TopologicalFrustrationOnTorus}c.
This gives evidence that topological frustration can be successfully embedded into doubly periodic structures.
As shown in Figure~\ref{fig:TopologicalFrustrationOnTorus}d, e, and~f, such species can be easily found among nanostructures:
carbon nanotubes, nanotori, and graphene nanomeshes (GNM).

Figure~\ref{fig:TopologicalFrustrationOnTorus}e presents a torus that is different from those in Figure~\ref{fig:TopologicalFrustrationOnTorus}a and~b, for which topological frustration is demonstrated. Hence, we need to explain how the original graph was constructed and how other similar structures can be generated with ease. For this we need to recall Tutte's~\cite{Tutte1947}, Berge's~\cite{Berge1957,Berge1958}, and Gallai-Edmonds'~\cite{Gallai1964,Edmonds1965} theorems from graph theory which totally demystify topological frustration. Tutte's theorem provides a general criterion for the perfect pairing existence. A perfect pairing, called a perfect matching in the graph, is impossible if there is a subset of points $S$, which once deleted from the graph leads to the number of odd subgraph components $D$ that is larger than size of the deleted subset. Berge further clarified this rule proving that when the difference between the number of odd subgraph components and size of the deleted subset attains maximum it is equal the deficit of the graph~\cite{Berge1958}: $|D|-|S| = \eta$, where $|...|$ denotes the size of a set. The subset on which deficit is reached is called \emph{obstruction set}. Basically the obstruction set is the reason why topological frustration exists and why the maximum possible pairing is not perfect. Actually, Berge provided also a criterion for a pairing to be a maximum matching. Berge's criterion is utilized by all maximum matching algorithms mentioned above. The criterion states that the matching is maximum if and only if there are no \emph{augmenting paths} in the graph: such paths that alternate matched and unmatched edges, while being started and ended at a free points not covered by the matching. The Gallai-Edmonds' theorem provides a way to build the canonical obstruction set. Here it is enough to mention only that such canonical obstructions can be computationally found with the blossom algorithm; see, for example, \emph{NanographenesBuilder} software~\cite{NanographenesBuilderSaroka}.
Armed with the above knowledge, we can readily notice that the graph deficit is preserved if any of the subgraph components is amended with an even number of points. Also, we notice that even subgraph components $C$ does not influence the deficit of the graph. Thus, having found any obstruction set (using some algorithm or even manually) one can extend the graph to a huge variety of shapes and forms, preserving its deficit with these so-called \emph{extension rules}.
To find an obstruction set, one basically must classify graph point vertices into $S$, $D$ and $C$ sets. When $S$ represents a canonical obstruction then the decomposition is referred to as the Gallai-Edmonds decomposition. In a more general case, when $S$ is not a canonical obstruction this decomposition is called the Tutte-Berge decomposition. 

\begin{figure}
\centering
  \includegraphics[width=0.75\linewidth]{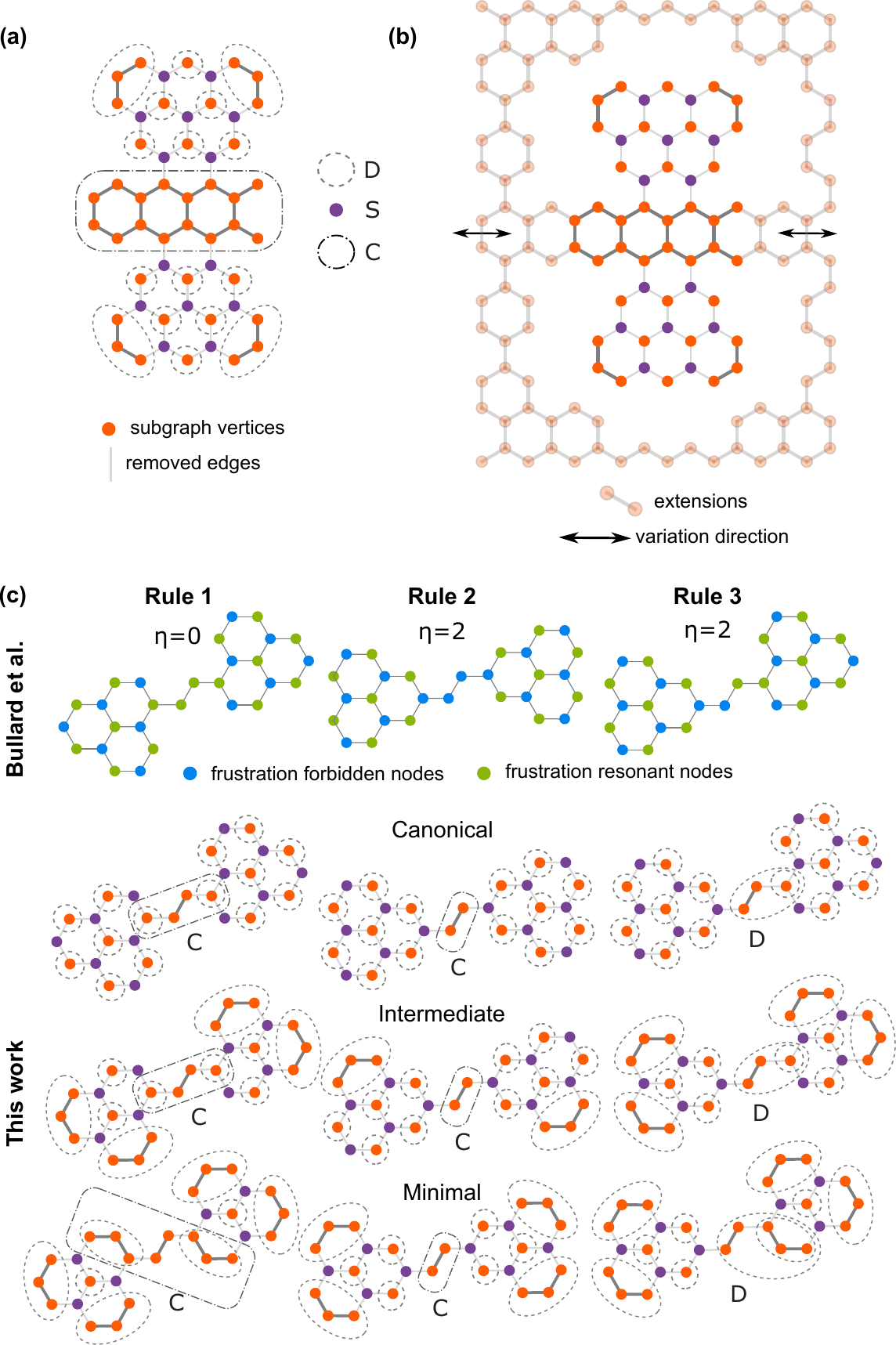}
  \caption{Obstructions and various rules. (a)  The Tutte-Berge decomposition of the graph from Ref.~\cite{Saroka2026b}, showing obstruction set $S$, together with even subgraph $C$ and odd subgraphs $D$. Edges incident on $S$ are removed edges depicted with reduced opacity. (b) The graph obtained from the graph in panel (a) by adding vertices and edges, which are both highlighted as semitransparent extensions. Black arrows: possible extension variations. (c) The Tutte-Berge decompositions for the Rules reported by Bullard et al. [adapted with permission from Ref.~\cite{Bullard2014}. Copyrighted by the American Physical Society.] 
  Canonical, minimal and intermediate obstructions for the final structures are shown. For Rule~1, the obstruction sets are related to the original elements to be connected, since the final obstruction vanishes. For Rules~2 and~3, minimal obstructions are \emph{quasi-}minimal: they can be reduced to $|S|=4$ at the left elements. For ``This work", notations are similar to panels (a) and (b).}
  \label{fig:ExtensionRules}
\end{figure}

Figure~\ref{fig:ExtensionRules} shows Tutte-Berge decompositions to which we applied the extension rules. The structural decomposition of the starting graph is presented in Figure~\ref{fig:ExtensionRules}a.
The even subgraph component $C$ of this graph is extended with an even number of vertices as shown in Figure~\ref{fig:ExtensionRules}b, to obtain a structure rollable into a torus. Further extension can be done to obtain torus in Figure~\ref{fig:TopologicalFrustrationOnTorus}e.

The above presented theorems shed the light on previous attempts to reveal organizational principles of topological frustration such as \emph{frustration annihilation rules} in Bullard et al.~\cite{Bullard2014}. 
In Figure~\ref{fig:ExtensionRules}c, we interpret in terms of $S$, $D$, and $C$ sets examples, given in Figure~2 of Bullard et al.~\cite{Bullard2014} as a demonstration of the \emph{frustration annihilation rules}. Rule~1 regarding the linkage of frustration resonance nodes reduces the deficit of the graph (we use our term deficit instead of the \emph{frustration unit} suggested by Bullard et al.) because it links two odd subgraph components, transforming them into a single even subgraph component $C$. This is a topologically frustrated structure of class~II. Rule~2 regarding the linkage of frustration forbidden nodes does not reduce the graph deficit because the link between two structures is an even subgraph component $C$, which cannot affect the graph deficit.
Rule~3 regarding the linkage of frustration-forbidden and frustration resonant nodes presents a structure that is of class~I. It features sublattice imbalance. This case is also covered by the theorems. Using an even number of vertices for the connector simply increases the odd subgraph component $D$.

We need to mention the flexibility of the Tutte-Berge decomposition. As shown in Figure~\ref{fig:ExtensionRules}c, the set $S$ can take different forms: canonical, intermediate and minimal. The latter was nicely formulated and discussed by David Sumner~\cite{Sumner1974,Sumner1976}, who found that vertices of a minimal obstruction set must be centers of claws, i.e. subgraphs representing complete bipartite graphs $K_{1,3}$. These obstructions are exemplified in Figure~\ref{fig:ExtensionRules}c in the row of minimal obstructions.

In a less technical way, we have seen that graph theory provides a comprehensive formalism and a \emph{reliable set of theorems} for tracking and handling topological frustration in bipartite graphs, including those composed of the integer number of hexagons of a honeycomb lattice. 
Specifically, topological frustration boils down to a collection of sites that form an obstruction for the lattice site pairwise coupling. This collection of sites stays invariant if the piece of lattice in question that forms the nanostructure is extended further into the lattice, keeping the even-odd parity balance with respect to structural components surrounding the obstruction. This freedom allows one to extend nearly any finite hydrocarbon into a two dimensional graphene nanomesh to obtain a fully flat band.

\subsection{A heuristic toy model\label{sec:PhysicalToyModel}}
The minimum example of topologically frustrated nanographene consists of $11$~hexagons and $38$ atoms -- Clar's goblet synthesized in 2020~\cite{Mishra2020a}. Extending such a structure into a periodic~\cite{Saroka2026b} or doubly periodic structure as in Figure~\ref{fig:ExtensionRules}~b further increases the size of the system, which makes it a bit challenging to grasp the origin of zero-energy states and their flatness with respect to the electron momentum $\vec{k}$. However, a heuristic toy model withing the single orbital ($p_z$) nearest-neighbor TB formalism can shed additional light on the subject. Below we construct such a model step-by-step starting from a simple pair of atoms and extend it into an alternating Su-Schrieffer-Heeger type model~\cite{Su1979,Su1980} that is to be associated with augmenting paths used in the matching theory as criterion for the maximum matching existence, following the theorem by Berge~\cite{Berge1957}.

Consider first a pair of two atoms carrying a single $p_z$ orbital each and no on-site energies. Let us assume that the orbitals are not hybridized and the atoms are not coupled, then the elementary Hamiltonian is a zero matrix leading to a pair of zero-energy states:
\begin{align}
    \left(\begin{array}{cc}
      0   & 0  \\
      0  & 0 
    \end{array}\right) & \rightarrow \epsilon_{1,2} = 0 \, .
    \label{eq:TwoAtomsUncoupled}
\end{align}
Next let us gradually switch on inter-site coupling $t$. The net outcome is the energy splitting:
\begin{align}
    \left(\begin{array}{cc}
      0   & t  \\
      t  & 0 
    \end{array}\right) & \rightarrow \epsilon_{1,2} = \pm t \, .
    \label{eq:TwoAtomsCoupled2}
\end{align}
Thus, we see that any interacting pair of orbitals results in the zero-energy level splitting.

At the next stage let us consider a longer chain consisting of four~atomic sites and four~orbitals numbered in ascending order from the left to the right. We skip an odd chain with three~atomic sites because topological frustration of class~II is specific for even number of sites and a quasi-symmetry that maintains equal number of sites in the bipartite sublattices. The latter is obviously the case for the four~atom chain. Similarly to Equation~(\ref{eq:TwoAtomsUncoupled}), four fully uncoupled atomic sites yield four~zero-energy levels. Introducing a coupling $t$ between the two atoms in the center results in
\begin{align}
    \left(\begin{array}{cccc}
      0  & 0  & 0 & 0 \\
      0  & 0  & t & 0 \\
      0  & t  & 0 & 0 \\
      0  & 0  & 0 & 0 \\
    \end{array}\right) & \rightarrow \epsilon_{2,3} = \pm t\, ; \qquad  \epsilon_{1,4} = 0\, \, .
    \label{eq:TwoAtomsCoupled3}
\end{align}
The net outcome is two zero-energy states and two non-zero energies resulting form the splitting due to $t$. Now notice that the central coupled pair of atoms is, in fact, a matching, while virtual (imaginable) links $1 \longleftrightarrow 2$ and $3 \longleftrightarrow 4$ can be denoted as regular graph edges. In doing so, one gets a path from site~$1$ to site~$4$ composed of alternating unmatched and matched edges. This is called an alternating path in the graph theory. Berge's theorem~\cite{Berge1957} suggests that if such a path starts and ends at free vertices not covered by the matching, then it is an \emph{augmenting path}, for which the matching size can be incremented by one via swapping matched and unmatched edges. Now that we have this knowledge, let us see what are the consequences of this abstract manipulation for our physical model. The exchanging of matched and unmatched edges correspond to introducing couplings between atoms~$1$ and~$2$, as well as between atoms~$3$ and~$4$. Note that now we have two couplings instead of one in the preceding case. The TB Hamiltonian and energy levels are
\begin{align}
    \left(\begin{array}{cccc}
      0  & t  & 0 & 0 \\
      t  & 0  & 0 & 0 \\
      0  & 0  & 0 & t \\
      0  & 0  & t & 0 \\
    \end{array}\right) & \rightarrow \epsilon_{1,2} = \pm t\, ; \qquad  \epsilon_{3,4} = \pm t\, \, .
    \label{eq:TwoPairsOfCoupledAtoms}
\end{align}
As one can notice, there are no zero-energy levels anymore. A sizable energy gap opens between the energy levels. For this case, we have obtained atoms to be fully pairwise coupled. Graph theory wise, this pairing corresponds to the perfect matching.

The linear chain, of course, is not sufficient to realize topological frustration. However, it provides a useful relation between the physical Su-Schrieffer-Heeger model and augmenting paths of the graph theory which can be used further. Now the key thing to understand is that when structure is topologically frustrated it does not allow augmenting paths~\cite{Berge1957} even though it can contain vertices not covered by matchings or physically speaking free standing uncoupled atoms. 
One can try to parametrize nearest-neighbor couplings corresponding to all matchings as $t_1 \propto  \tau$, while the remaining couplings as $t_2 \propto 1/\tau$ and see that by varying $\tau$ from~$0$ to~$\infty$, it is impossible to open the gap in the system. This persists for any imaginable nearest-neighbor coupling configuration of the given type.
Thus, one can view the zero-energy modes arising form topological frustration as a mathematical combinatorial magic in a sense. The other theorems and decompositions presented in the previous section, provide a nicely organized way for a reliable navigation through the huge variety of of honeycomb lattice structures that preserve topological frustration. The actual physical ingredient playing important role here is the chiral symmetry that maintains bipartite structure of the lattice and ensures vanishing on-site energies, i.e. diagonal elements of the Hamiltonian.

The presented here heuristics relies only on limiting cases $\tau \rightarrow 0$ and $\tau \rightarrow \infty$. The rigorous mathematical proof is supplied by the nullity theorem of Fajtlowicz et al.~\cite{Fajtlowicz2005}, which is a good supplement for the simple model above.

Regarding the TB model, we note that, for periodic structures, the bare $t$ couplings can be replaced by functions of the form: $f_{\vec{k}} = \sum_j t_j \exp \left(i \vec{k} \cdot \vec{a}_j \right)$, where $\vec{a}_j$ are the vectors to the nearest neighbors. Whence it follows that vanishing $t$'s eliminate $\vec{k}$-dependence altogether.

It is also worth noting that, in general, the TB Hamiltonian of the topologically frustrated system can be dropped into the form of Equation~3 in Ref.~\cite{Calugaru2022}. By the K{\"o}nig's theorem~\cite{Konig1931}, in a bipartite graph, the maximum matching size is equal to the size of a minimum vertex cover. Therefore, taking the minimum vertex cover and a maximum independent vertex set, sum of which is equal to the total number of vertices in the graph, as two partition groups, 
puts the TB Hamiltonian directly into the form of Equation~3 in Ref.~\cite{Calugaru2022}, where the maximum independent vertex set block will be the one proportional to the identity~\cite{NoteBernevig2026}.
The difference between the size of the maximum independent vertex set and the size of the minimum vertex cover provides the number of guaranteed flat bands according to the general scheme of Ref.~\cite{Calugaru2022}. We shall note that Fajtlovich et al.~\cite{Fajtlowicz2005} nullity theorem for a bipartite case defines the number of zero eigenvalues in the bipartite graph spectrum as not less than the difference between  the size of the maximum independent vertex set and the size of the maximum matching. Both just defined differences are, in fact, equal to the graph deficit $\eta$.

The above-drawn conclusions are corroborated with the actual numerical modeling in the next section.

\subsection{Flat bands and emergent quantum magnetism\label{sec:FlatbandsAndMagnetism}}
In the previous Section~\ref{sec:GraphTheory}, we have seen that the GNM shown in Figure~\ref{fig:TopologicalFrustrationOnTorus}f is characterized by topological frustration. From the quantum chemistry point of view, this means that the structure shall exhibit properties of a concealed non-Kekule diradical. The presence of two zero eigenvalues in the adjacency matrix of the graph in Figure~\ref{fig:TopologicalFrustrationOnTorus}a also supports this idea. Then it is quite natural to ponder the emergence of magnetism and magnetic excitations. 
Therefore, we proceed to a direct verification using honest, as compared to the heuristics in Section~\ref{sec:PhysicalToyModel}, the nearest-neighbor TB model and a further inspection of magnetism with the TB+U, which can be mapped to the effective Heisenberg spin lattice for studying magnetic excitations. The Hamiltonian, which encompasses both models, is~\cite{FernandezRossier2007,FernandezRossier2008,Claveau2014}
\begin{equation}
   H= \sum_{i,\sigma} \varepsilon \hat{c}^{\dagger}_{i,\sigma} \hat{c}_{i,\sigma} + \sum_{<i,j>, \sigma} t_{1} \left( \hat{c}^{\dagger}_{i,\sigma} \hat{c}_{j,\sigma} + \hat{c}^{\dagger}_{j,\sigma} \hat{c}_{i,\sigma}\right) + U \sum_i \hat{n}_{i,\uparrow} \hat{n}_{i,\downarrow}
  \label{eqn:TBHamiltonian}
\end{equation}
where $\hat{c}^{\dagger}_{i,\sigma}$ and $\hat{c}_{i,\sigma}$ are creation and annihilation operators of an electron with spin $\sigma = \uparrow,\downarrow$ at site $i$, respectively, $\hat{n}_{i,\sigma} = \hat{c}^{\dagger}_{i,\sigma} \hat{c}_{i,\sigma}$ is the number operator for an electron with spin $\sigma$ at site $i$, $\varepsilon$ is the on-site energy that can be set to zero for all lattice sites and spin directions without loss of generality ($\varepsilon=0$), $t_{1}=3.12$~eV~\cite{Partoens2006} is the nearest-neighbor hopping integral. We note here that $t_1$ varies for different structures~\cite{Saroka2018a,Payod2020}, therefore, in what follows, we will use dimensionless energy units,  $E/t_1$, where $E$ is the eigenstate energy. $U$ is the on-site Coulomb repulsion. $U/t_1 \sim 1.3$ usually gives good agreement with the generalized gradient approximation to density functional theory~\cite{Yazyev2008} (note that slightly different values,  such as $U/t_1 \sim 1.6$ as an effective interaction, have also been reported~\cite{Schuler2013}). In the mean-field approximation the $3$-rd term of $H$ in Equation~(\ref{eqn:TBHamiltonian}) reduces to 
\begin{equation}
    U \sum_i \left( \hat{n}_{i,\uparrow} \langle \hat{n}_{i,\downarrow} \rangle + \hat{n}_{i,\downarrow} \langle \hat{n}_{i,\uparrow} \rangle - \langle \hat{n}_{i,\uparrow} \rangle \langle \hat{n}_{i,\downarrow} \rangle \right) \, ,
    \label{eq:MeanFieldTermHubbard}
\end{equation}
where $\langle \ldots \rangle$ stands for an average, leading to a coupled spin-up and -down eigenproblems to be solved self-consistently via iterations, starting from some initial assumption. The problem is solved using \emph{TBpack} package~\cite{TBpackSaroka}. The $32 \times 32$ $\vec{k}$-grid is used in TB and TB+U calculations. In the TB+U calculations, the occupation numbers convergence threshold and smearing temperature are set as $10^{-5}$ and $T=30$~K, respectively.

\begin{figure}
\centering
  \includegraphics[width=0.90\linewidth]{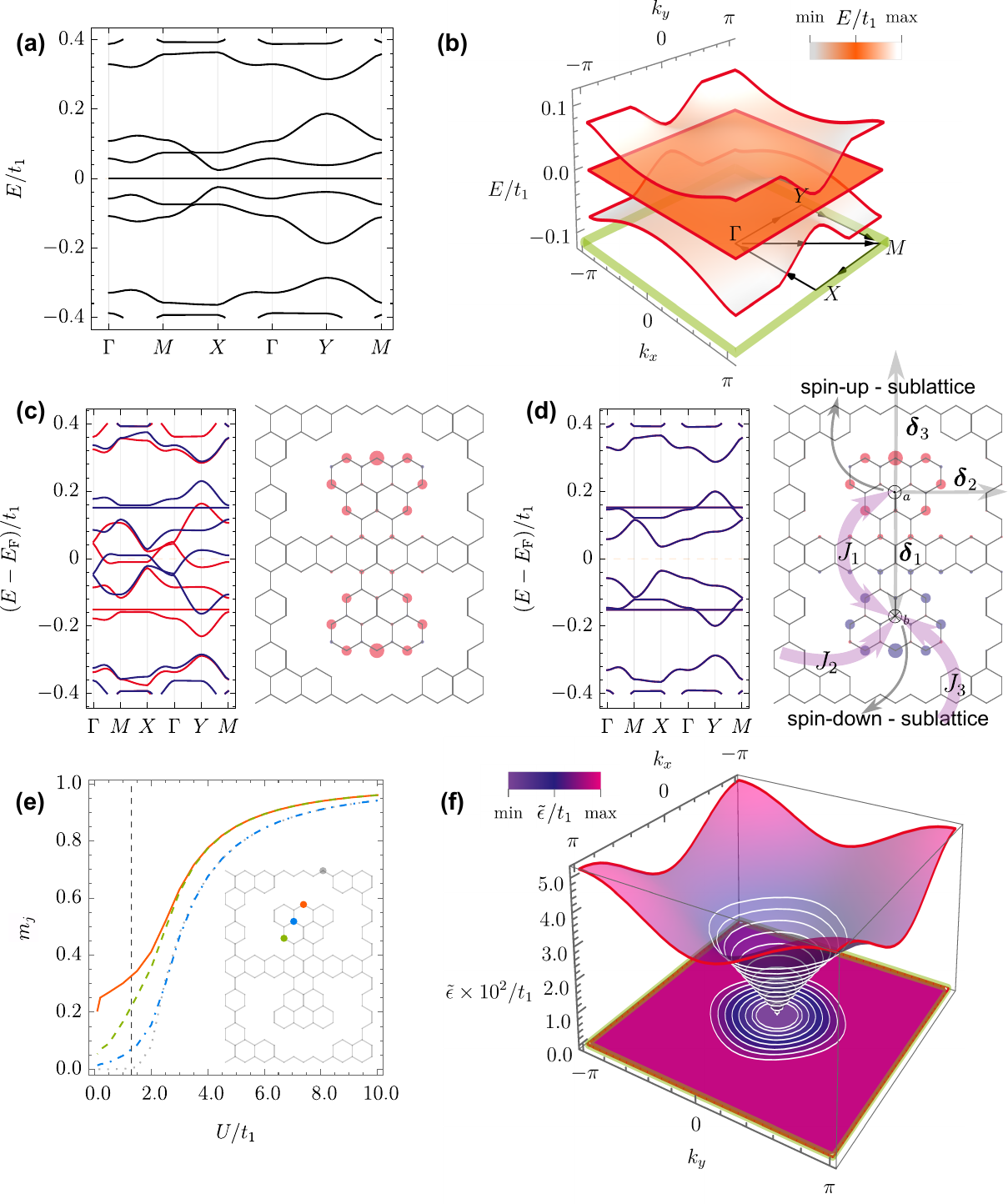}
  \caption{Flat bands and magnetism. (a) The TB energy bands along a 1D $k$-path through the high symmetry points. (b) The 2D plot of the four TB bands closest to the Fermi level. Light green: Brillouin zone. Black arrows: the $k$-path through the high symmetry points.  (c) The TB+U band structure (left) and the spin density (right) for the initial FM spin configuration. Spin-up and -down species are represented by red and blue colors, accordingly. Thin dashed orange line marks the Fermi level adjusted to zero. (d) Same as panel (c) but for the initial AFM spin configuration. $\odot_a$ and $\otimes_b$ mark positions of the positive and negative magnetization gravity centers, respectively. Semitransparent magenta arrows denote exchange interactions $J_1$, $J_2$, and $J_3$. Semitransparent gray arrows denote effective Heisenberg model nearest neighbor vectors $\vec{\delta}_1$, $\vec{\delta}_2$, and $\vec{\delta}_3$. (e) The magnetization of the four lattice sites, highlighted within the unit cell of the GNM in the inset as green, blue, orange, and gray dots, versus $U/t_{1}$. Dashed gray vertical line denotes the $U/t_{1}$-value used in panels~(c) and~(d). (f) The magnon dispersion from Equation~(\ref{eq:MagnonDispersion}). $\tilde{\epsilon}=0$ plane of the 3D plot hosts a 2D contour plot of the dispersion with the same color scheme. Light green: Brillouin zone. White: iso-energy curves.}
  \label{fig:FlatbandsAndMagnetism}
\end{figure}

Figure~\ref{fig:FlatbandsAndMagnetism} displays the electronic properties of the GNM in question. The TB bands plotted along a $\vec{k}$-path through the high symmetry points presented in Figure~\ref{fig:FlatbandsAndMagnetism}a contain two flat bands at $E_{\mathrm{F}}=0$~eV. As can be seen in Figure~\ref{fig:FlatbandsAndMagnetism}b, the flat bands are truly flat throughout the Brillouin zone. The TB+U calculations in Figure~\ref{fig:FlatbandsAndMagnetism}c, d for the FM and AFM initial spin configurations show spin-polarizations, as expected from the Stoner-Hubbard criterion~\cite{Stoner1938,Stoner1939,Hubbard1963}: $U\rho \gg 1$, where $\rho$ is the density of states at the Fermi level. The AFM configuration has a lower energy compared to the FM configuration by $0.053 t_1$. Thus, in agreement with Lieb's theorem~\cite{Lieb1989} and Ovchinnikov's rule~\cite{Ovchinnikov1978}, both stating that the ground state spin configuration of the Hubbard model on a bipartite lattice is  $S= \left|N_A - N_B\right|/2$, where $N_{A/B}$ is the number of atoms in the sublattice, the AFM configuration is the ground state of the system. The atomic site magnetization, shown in Figure~\ref{fig:FlatbandsAndMagnetism}e, demonstrates that non-zero magnetization persists till $\frac{U}{t_1}\rightarrow 0$. The largest on-site magnetization at $\frac{U}{t_1}\rightarrow 0$  exceeds $0.2$ for zigzag graphene nanoribbons~\cite{Fujita1996}.

The spatially localized and spin-polarized electronic modes, presented in Figure~\ref{fig:FlatbandsAndMagnetism}d, can effectively act as a single particle spins placed at points representing the average positions of the atomic magnetizations: \begin{align}
\vec{R}_{\uparrow/\downarrow} &= \dfrac{\sum_{j=1}^{N_{\uparrow/\downarrow}} m_{j} \vec{r}_j}{\sum_{j=1}^{N_{\uparrow/\downarrow}} m_{j}}\,, 
\label{eq:SpinGravityCenters}
\end{align}
where $m_{j} = \left(n_{\uparrow,j}-n_{\downarrow,j}\right)/2$ and $r_j$ are $j$-th site magnetization (i.e. spin density defined via spin-up and -down occupation numbers $n_{\uparrow/\downarrow,j}$) and position, respectively, $N_{\uparrow/\downarrow}$ means summation over sites with either positive (spin-up $\uparrow$) or negative (spin-down $\downarrow$) magnetizations. The effective gravity centers of magnetization $\vec{R}_{\uparrow/\downarrow}$ form a superlattice over the nanoscale honeycomb lattice. As follows from Figure~\ref{fig:FlatbandsAndMagnetism}d, this shall be a rectangular spin lattice with one spin-up and one spin-down per unit cell. In what follows, we derive the dispersion relation for magnetic excitations in this effective spin superlattice. The spin lattice can be described by Heisenberg Hamiltonian, which exchange couplings can be estimated from the TB+U model total energies.

The general spin lattice built on the magnetization gravity centers presented in Figure~\ref{fig:FlatbandsAndMagnetism}d can be described by Heisenberg exchange Hamiltonian similar to other lattices such as the one in yttrium iron garnet~\cite{Cherepanov1993,Shen2018}. The model Hamiltonian reads
\begin{align}
    \Cal{H} &= -\dfrac{1}{2} \sum^{N}_{n=1} J_{aa} \vec{S}_a (\vec{r}_{na}) \sum_{\vec{\delta}_2} \vec{S}_a(\vec{r}_{na}+\vec{\delta}_2) - \nonumber \\
    &\phantom{=}- \dfrac{1}{2} \sum^{N}_{n=1} J_{bb} \vec{S}_b (\vec{r}_{nb}) \sum_{\vec{\delta}_2} \vec{S}_b(\vec{r}_{nb}+\vec{\delta}_2) - \nonumber \\
    &\phantom{=}-
    \sum^{N}_{n=1} \vec{S}_a (\vec{r}_{na}) \left[J_{ab} \vec{S}_b(\vec{r}_{na}+\vec{\delta}_1) + J^{\prime}_{ab} \vec{S}_b(\vec{r}_{na}+\vec{\delta}_3)\right]\, ,
    \label{eq:HeisenbergHamiltonian}
\end{align}
where $a$ and $b$ denote the two sublattices of the spin lattice (two spins in the unit cell of the structure, $S_a$ and $S_b$, are generally treated as different in magnitude for a while):
\begin{align}
  \vdots \; \; \: \qquad \qquad \qquad \nonumber \\
    \ldots \quad \begin{array}{|c|c|c|c|}
    \hline
       \odot_a   & \odot_a & \odot_a & \odot_a \\
       \otimes_b & \otimes_b & \otimes_b& \otimes_b  \\
       \hline
        \odot_a & \odot_a & \odot_a & \odot_a \\
       \otimes_b& \otimes _b& \otimes_b& \otimes_b \\
       \hline
    \end{array} \quad \ldots \nonumber \\
    \vdots \; \: \qquad \qquad \qquad \, , \nonumber 
\end{align}
where $\otimes$ and $\odot$ symbols denote spin-down and -up, respectively, $J$'s stand for the exchange couplings between nearest-neighbor spins of the corresponding sublattices, $^\prime$ is used to distinguish intercellular coupling from the intracellular one, where confusion can arise, $\vec{\delta}_i$'s denote directional vectors to the nearest neighbor spins: $1$ - intracellular vector pointing down, $2$ - intercellular horizontal vector pointing right, $3$ - intercellular vertical vector pointing up; see also Figure~\ref{fig:FlatbandsAndMagnetism}d. $N$ is the number of the unit cells in the spin lattice. By switching to ladder operators $S^{\pm} = S_x \pm i S_y$, one can introduce a linearized Holstein-Primakoff transform~\cite{Holstein1940}:
\begin{align}
    S^z_a &= S_a - \Goth{a}^{\dagger} \Goth{a}  & S^{+}_a & = \sqrt{2 S_a - \Goth{a}^{\dagger} \Goth{a}} \; \Goth{a} \approx \sqrt{2 S_a}\; \Goth{a} \nonumber \\
    S^z_b &=  \Goth{b}^{\dagger} \Goth{b} - S_b & S^{+}_b & = \Goth{b}^{\dagger} \sqrt{2 S_b - \Goth{b}^{\dagger} \Goth{b}} \approx \Goth{b}^{\dagger} \sqrt{2 S_b}  \, ,
    \label{eq:HolsteinPrimakoff}
\end{align}
where $S^{-}_{a,b} = \left(S^{+}_{a,b}\right)^{\dagger}$, and rewrite Hamiltonian in Equation~(\ref{eq:HeisenbergHamiltonian}) in terms of $\Goth{a} (\Goth{a}^{\dagger})$ and $\Goth{b} (\Goth{b}^{\dagger})$, which are the annihilation (creation) operators for the bosonic magnon excitations on sublattices $a$ and $b$, respectively. Note that Equations~(\ref{eq:HolsteinPrimakoff}) accounts for AFM ordering on sublattices $a$ and~$b$. Then transforming thus-obtained Hamiltonian into $\vec{k}$-space by Fourier transforms:
\begin{align}
    \Goth{a}_{nj} &= \dfrac{1}{\sqrt{N}} \sum_{\vec{k}} \Goth{a}_\vec{k}  e^{\mathrm{i} \vec{k} \cdot \vec{r}_{nj} };& 
\Goth{b}_{nj} &= \dfrac{1}{\sqrt{N}}  \sum_{\vec{k}} \Goth{b}_\vec{k}  e^{\mathrm{i} \vec{k} \cdot \vec{r}_{nj} } \, ,
\end{align}
where $j=a,b$ stands for the lattice site, we get a tight-binding-like Bogoliubov-de Gennes Hamiltonian:
\begin{align}
    \Cal{H}(\vec{k}) &= \Cal{H}_0 + \Cal{H}_1 (\vec{k}) \label{eq:HeisenbergHamiltonianK} \\
    \Cal{H}_0 &= N \left[\left(J_{ab} + J^{\prime}_{ab}\right) S_a S_b - J_{aa} S^2_a - J_{bb} S^2_b\right]\label{eq:IsingEnergy}\\
    \Cal{H}_1(\vec{k}) &= \sum_{\vec{k}} A(\vec{k}) \Goth{a}^{\dagger}_{\vec{k}} \Goth{a}_{\vec{k}} + \sum_{\vec{k}} B(\vec{k}) \Goth{b}^{\dagger}_{\vec{k}} \Goth{b}_{\vec{k}} + \left[\sum_{\vec{k}} C(\vec{k}) \Goth{a}_{\vec{k}} \Goth{b}_{\vec{-k}} + \text{h.c}\right]
\end{align}
where $\Cal{H}_0$ is the classical (Ising) ground state energy and $\Cal{H}_1(\vec{k})$ is the magnon dispersion and
\begin{align}
    A(\vec{k}) &= 2 J_{aa} S_a \left[1 - \cos \left(\vec{k}\vec{\delta}_2\right)\right] - \left(J_{ab} + J^{\prime}_{ab}\right) S_b\\
    B(\vec{k}) &= 2 J_{bb} S_b \left[1 - \cos \left(\vec{k}\vec{\delta}_2\right)\right] - \left(J_{ab} + J^{\prime}_{ab}\right) S_a \\
    C(\vec{k}) &= - \sqrt{S_a S_b} \left(J_{ab} e^{-\mathrm{i} \vec{k} \vec{\delta}_1} + J^{\prime}_{ab} e^{-\mathrm{i} \vec{k} \vec{\delta}_3}\right) \, .
\end{align}
Given that $A^{\ast}(\vec{k}) = A(-\vec{k})$, $B^{\ast}(\vec{k}) = B(-\vec{k})$, and $C^{\ast}(\vec{k}) = C(-\vec{k})$. Bosonic diagonalization of $\Cal{H}_1$, which relies on a paraunitary transform~\cite{VanHemmen1980}, is equivalent to the standard eigenvalue problem:
\begin{equation}
    \left(\begin{array}{cc}
        A & C \\
        -C^{\dagger} & -B
    \end{array}\right) \Psi = \epsilon \Psi \, ,
\end{equation}
which gives half of the eigenvalues of the full bosonic Hamiltonian:
\begin{equation}
    \epsilon_{1,2} = \dfrac{(A-B)\pm \sqrt{(A+B)^2 - 4 |C|^2}}{2}
\end{equation}
The remaining eigenvalues can be obtained by negating the eigenvalues $\epsilon$.

The general model in Equation~(\ref{eq:HeisenbergHamiltonianK}) can be specialized now to our case by setting $J_{ab} \equiv J_1$, $J^{\prime}_{ab} \equiv J_3$, $J_{aa} = J_{bb} \equiv J_2$ (see Figure~\ref{fig:FlatbandsAndMagnetism}d), $S_a = S_b \equiv S$. Thence, we have 
\begin{align}
    \tilde{\epsilon}_{1,2} &= \pm \sqrt{\Cal{A}^2(\vec{k}) - |\Cal{C}(\vec{k})|^2} \\
    \Cal{A}(\vec{k}) &= \left\{2 J_2 \left[1 - \cos\left(\vec{k} \vec{\delta}_2\right)\right] - J_1 - J_3\right\} S \\
    \Cal{C}(\vec{k}) &= - \left(J_1 e^{-\mathrm{i} \vec{k} \vec{\delta}_1} + J_3 e^{-\mathrm{i} \vec{k} \vec{\delta}_3}\right) S
\end{align}
or, equivalently,
\begin{equation}
    \tilde{\epsilon}_{1,2} = \pm S \sqrt{\left\{J_1 + J_3 - 2 J_2 \left[1 - \cos \left(\vec{k} \vec{\delta}_2\right)\right]\right\}^2 - J_1^2 - J^2_3 - 2 J_1 J_3 \cos \left[\vec{k} \left(\vec{\delta}_1 - \vec{\delta}_3\right)\right]}\, .\label{eq:MagnonDispersion}
\end{equation}
The dispersion relation in Equation~(\ref{eq:MagnonDispersion}), comprises two important limiting cases. If $J_1=J_3=0$ then $\tilde{\epsilon}_{1,2} = \pm 2 J_2 S \left[1 - \cos \left(\vec{k} \vec{\delta}_2\right)\right]$, which is a typical 1D FM magnon dispersion. On the other hand, if $J_2 = 0$, $J_3 = J_1$, and $\vec{\delta}_3 = - \vec{\delta}_1$, Equation~(\ref{eq:MagnonDispersion}) yields $\tilde{\epsilon}_{1,2} = \pm 2 |J_1| S \left|\sin\left(\vec{k} \vec{\delta}_1\right)\right|$, wherein the 1D AFM magnon dispersion is readily recognized.
Thus, the topologically frustrated GNM in question possesses anisotropic magnon dispersion combining FM and AFM magnons. 

The exchange couplings $J_1$, $J_2$ and $J_3$ can be estimated by relating the classical ground state energies per unit cell $\Cal{H}_0/N$, see Equation~(\ref{eq:IsingEnergy}), to the total energies obtained within the TB+U model, similar to what is done Refs.~\cite{Jiang2021,Anindya2026,Catarina2023,Henriques2024,Phillips2026}. For the four configurations:
\begin{align*}
    \begin{array}{|c|c|c|c|}
    \hline
       \odot_1 & \odot & \odot & \odot \\
       \otimes_2& \otimes_3 & \otimes& \otimes  \\
    \hline
        \odot_4 & \odot & \odot & \odot \\
       \otimes& \otimes & \otimes& \otimes\\
    \hline
    \end{array} & 
    &
    \begin{array}{|c|c|c|c|}
    \hline
       \odot_1  & \odot & \odot & \odot \\
       \odot_2  & \odot_3 & \odot & \odot \\
    \hline
       \odot_4  & \odot & \odot & \odot \\
       \odot  & \odot & \odot & \odot \\
    \hline
    \end{array}
    & 
    &
    \begin{array}{|c|c|c|c|}
    \hline
       \odot_1  & \odot & \odot & \odot \\
       \otimes_2& \otimes_3 & \otimes& \otimes  \\
       \otimes_4& \otimes & \otimes& \otimes\\
       \odot  & \odot & \odot & \odot \\
    \hline
    \end{array}
    & &
        \begin{array}{|cc|cc|}
        \hline
       \odot_1  & \otimes & \odot & \otimes \\
       \otimes_2& \odot_3 & \otimes&  \odot \\
       \hline
       \odot_4  & \otimes & \odot & \otimes \\
       \otimes& \odot & \otimes&  \odot \\
       \hline
    \end{array}
\end{align*}
where our exchange couplings are obviously $J_1 = J_{12}$, $J_2=J_{23}$, and $J_3=J_{24}$ if $J_{ij}$ denotes a coupling between spins indexed with subscripts $i$ and $j$ in the schemes above (cf. Figure~\ref{fig:FlatbandsAndMagnetism}d), one has (in order from left to right) 
\begin{align}
    J_1 S^2 + J_3 S^2 - 2 J_2 S^2 &= E_1 - E_0 \,, \nonumber\\
    -J_1 S^2 - J_3 S^2 - 2 J_2 S^2 &= E_2 - E_0 \,, \nonumber\\
    J_1 S^2 - J_3 S^2 - 2 J_2 S^2 &= E_3 - E_0 \,, \nonumber\\
    J_1 S^2 + J_3 S^2 + 2 J_2 S^2 &= E_4 - E_0 \,, 
    \label{eq:Jfitting}
\end{align}
where $E_0$ is the spin-independent part of a total energy, $E_{1-4}$ are the total energies per two-spins unit cell extracted from TB+U model (four-spins unit cell energies are divided by $2$ to obtain $E_3$ and $E_4$). Note that $E_0$ is the fourth unknown along with $J_1$, $J_2$, and $J_3$ in the above set of simultaneous Equations~(\ref{eq:Jfitting}), therefore, it has a unique solution. $S=1/2$ is the spin of an unpaired electron.
Here we neglect fluctuations that are known to lower the classical energy of AFM configurations with respect to FM ones~\cite{Anderson1952}. The nearest-neighbor vectors $\vec{\delta}_{1,2,3}$ can be found by setting spin positions as gravity centers for positive and negative on-site magnetizations, as shown in Figure~\ref{fig:FlatbandsAndMagnetism}d.

Figure~\ref{fig:FlatbandsAndMagnetism}f presents the positive branch of the magnon dispersion in Equation~(\ref{eq:MagnonDispersion}) with the parameters given in Table~\ref{tab:SpinLatticeParameters}.
\begin{table}
\centering
 \caption{The summary of the spin lattice energy (in terms of $t_1$) and structural parameters.}
  \begin{tabular}[htbp]{ccccccccc}
    \hline
     $J_1$ & $J_2$ & $J_3$ & $E_0$ & $E_1$ (AFM) & $E_2$ (FM) & $E_3$ & $E_4$ & $\delta_1/\delta_3$ \\
    \hline
    $-0.0914$ & $0.0061$ & $-0.0142$ & $-176.949$ & $-176.978$ & $-176.925$ & $-176.971$ & $-176.972$ & $0.65$  \\
    \hline
  \end{tabular}
  \label{tab:SpinLatticeParameters}
\end{table}
Although some anisotropy is clearly observable, the magnon dispersion is AFM in nature with $\sin(\vec{k}\vec{\delta})\sim k$ at the $\Gamma$ point, as indicated by a Dirac cone. This characteristic can be attributed to the gentle $J_2$ exchange coupling. By setting $t_1\approx 3$~eV and contrasting the exchange couplings with the room temperature energy ($26$~meV) and Landauer limit for the minimum energy dissipation per bit ($k_{\mathrm{B}} T \ln(2) \approx 18$~meV, where $k_{\mathrm{B}}=8.617\times 10^{-2}$~meV/K is the Boltzmann constant, $T=300$~K is the room temperature)~\cite{Landauer1961}, one sees that the found couplings, even $J_2$, are large enough to promise near room-temperature spintronic operations.
While our values can be slightly overestimated due to the simplicity of the model used, they are well within the range of experimental values reported for the topologically frustrated Clar's goblet ($23$~meV)~\cite{Mishra2020a} and in agreement with those reported for triangulene-based covalent organic frameworks~\cite{Phillips2026}. For comparison, the largest exchange coupling in yttrium iron garnet is about $6$~meV~\cite{Princep2017}.
The found coupling energies also belong to the terahertz frequency energy range ($4$~meV $\propto 1$~THz), which means that the spin-wave excitations must be responsive and exhibit ultrafast dynamics, thus adding topologically frustrated GNM magnonics to the pool of candidates to bridge the terahertz gap~\cite{Hartmann2014}. 
Although such meshes can be challenging for synthesis and maintenance due to the high reactivity of zigzag (acene) edges, they could be stabilized upon encapsulation between two hexagonal boron nitride monolayers for ambient environment operation. Alternatively, hexagonal boron nitride patches could be used as fillers for vacancy regions~\cite{Wang2021a,Tepliakov2023b} or nanopatterning with hydrogen or fluorine~\cite{Chernozatonskii2010,Withers2011} could be used to reinforce the structure and design the required topologically frustrated $\pi$-electron networks. Thus, in general, such structures shall be deemed realistic and promising for practical applications.

\subsection{Other experimental realizations}
In the past decade, we have seen an intensive exchange of ideas between nanotechnology and condensed matter on one side and metamaterials on the other. Many effects, including topological ones, have found analogs in trapped atoms~\cite{Jotzu2014}, optical~\cite{Plotnik2014,Liu2021,Xia2023}, polaritonic~\cite{Milicevic2015}, microwave~\cite{Bellec2014}, electrical~\cite{Lee2018a}, acoustic~\cite{Peri2019}, and nanomechanical~\cite{Xi2021} systems. In this respect, it is worth noting that the high reactivity of acene edges mentioned at the end of Section~\ref{sec:FlatbandsAndMagnetism}, is not a problem for such artificially tailored and manipulated systems. Such systems, while lacking condensed matter effects, magnetism and superconductivity, can nevertheless provide a proof-of-concept for the flat band engineering based on topological frustration in 2D systems. The cold atom platform, for example, has demonstrated a coherent few thousand atoms manipulation~\cite{Manetsch2025} along with intricate interatomic couplings~\cite{Jotzu2014}, which are much more complex than the nearest neighbor couplings required for the topological frustration.

Regarding synthetic chemistry realizations, we point out that
several topologically frustrated hydrocarbons have been synthesized~\cite{Mishra2020a,Song2024,Jiao2025,Mishra2025,Imran2025,Li2026}.
Three new species emerged in last year only~\cite{Jiao2025,Mishra2025,Imran2025} and one more this year~\cite{Li2026}. Each of these new topologically frustrated molecules can be a seed for a particular topologically frustrated GNM.
The acene hydrocarbons, which may be a base element of such GNMs, have experimentally reached pentadecacene size~\cite{Jancarik2021,Jancarik2022,Ruan2024,Ruan2025}. This is three times longer than the base acene used in our model structure. As our graph theoretical consideration suggests, the acene bases can be further expanded into wider zigzag ribbon bases without losing topological frustration property. Zigzag graphene nanoribbons have been experimentally synthesize a decade ago~\cite{Ruffieux2016}. They can reach $10$~nm in length and be $6$ zigzag chains wide. Thus, other GNM configurations are certainly expected to be possible, provided we keep the current trends and momentum in the field of synthetic chemistry.

\section{Conclusion}
In summary, it has been demonstrated that graph-theoretic topological frustration can persist on a torus. This gives rise not only to zero-energy states in toroidal structures such as nanotori but also to flat energy bands at the Fermi level in 1D nanotubes and 2D nanomeshes. The latter exhibits emergent metal-free quantum magnetism, which is promising for tailoring AFM spintronics devices with ultra-fast near room temperature operation. Thus, topological frustration provides a tool for designing and systematically studying strongly correlated exotic phenomena in monolayer graphene, avoiding twistronics with bi-and tri-layers or multilayer stackings in few-layer systems. Since electronic flat bands are expected to result in an elevated superconductivity transition temperature~\cite{Khodel1990,Volovik1994,Kopnin2011,Kopnin2013,Peotta2015,Kauppila2016,Volovik2018,Peltonen2020,Nunes2020}, the superconducting phase diagram can be a subject of future work for such topologically frustrated systems. A totally different but equally appealing can be a research line on the topological frustration in quantum spin liquids and dimerized magnets, i.e. in systems, where each site (atom) of the initial lattice carries an intrinsic spin.

\medskip
\textbf{Supporting Information} \par 
Supporting Information is available from from the author.

\medskip
\textbf{Acknowledgements} \par 
The author thank C.~A.~Downing, G.~E.~Volovik, V.~R.~Shaginyan, J.~Jiang, V.~Lauranenka, D.~Corona, A.~Kamra, J.~Fern{\'a}ndez-Rossier and R.~Fasel for useful comments and discussions. V.A.S. was supported by the BRFFR Scientist-2026 (Grant No. F26U-002) and partly by HORIZON EUROPE MSCA-2021-PF-01 (project no. 101065500, TeraExc) for TBpack development.
\medskip

%
\bibliographystyle{MSP}
\bibliography{library.bib}

\begin{figure}
\centering
\textbf{Table of Contents}\\
\medskip
  \includegraphics{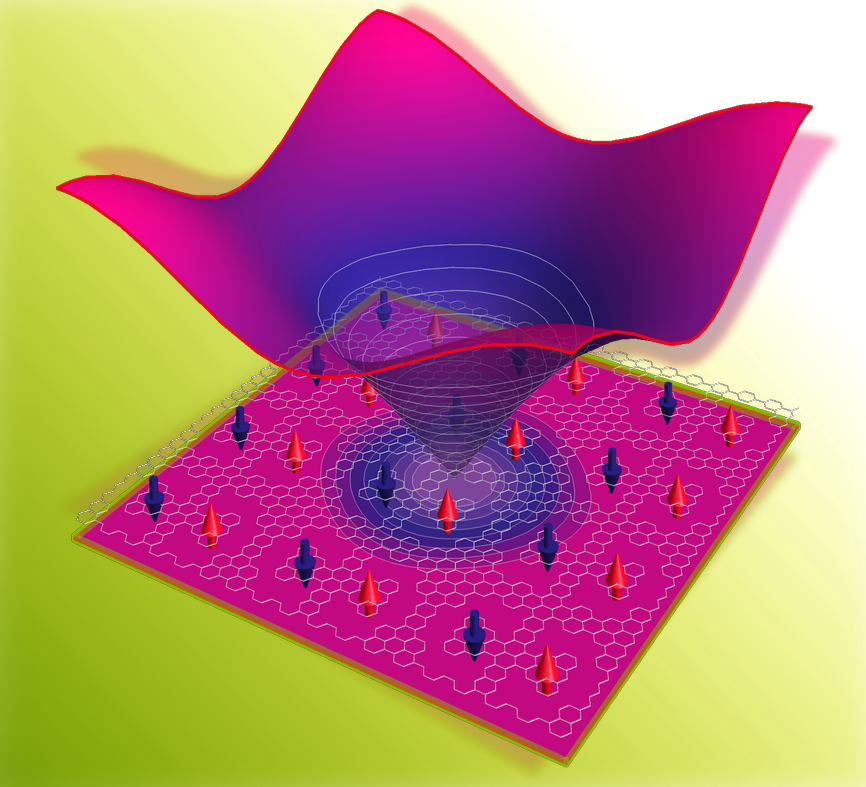}
  \medskip
  \caption*{Flat electronic bands are desired features in two dimensional condensed matter systems because they lead to emergent magnetism. In the bipartite honeycomb lattice with balanced sublattices, topological frustration can lead to flat bands and antiferromagnetism. Topological frustration can be designed in graphene nanomeshes providing a route to antiferromagnetic magnon dispersion.}
\end{figure}

\end{document}